\documentclass[%
notitlepage,%
onecolumn,%
floats,%
aps,%
prd,%
nobibnotes,%
nofootinbib,%
amsmath,%
amssymb,%
amsfonts,%
amscd,%
superscriptaddress,%
eqsecnum%
]{revtex4-1}

\usepackage[utf8]{inputenc}
\usepackage{graphicx, array, dcolumn}
\usepackage[paperwidth=210mm, paperheight=297mm,
top=24mm,bottom=24mm,inner=20mm,outer=20mm,marginparwidth=20mm]{geometry}
\usepackage{subcaption}

\begin{document}

\title{Evolution of thick domain walls in inflationary and $p=w\rho$ universe}

\author{A.D.\,Dolgov}
\email{dolgov@fe.infn.it}
\affiliation{Novosibirsk State University, Novosibirsk, 630090, Russia}
\affiliation{Institute for Theoretical and Experimental Physics, Moscow, 117218, Russia}

\author{S.I.\,Godunov}
\email{sgodunov@itep.ru}
\affiliation{Novosibirsk State University, Novosibirsk, 630090, Russia}
\affiliation{Institute for Theoretical and Experimental Physics, Moscow, 117218, Russia}

\author{A.S.\,Rudenko}
\email{a.s.rudenko@inp.nsk.su}
\affiliation{Novosibirsk State University, Novosibirsk, 630090, Russia}
\affiliation{Budker Institute of Nuclear Physics, Novosibirsk, 630090, Russia}

\begin{abstract}
  We study the evolution of thick domain walls in the different models of cosmological inflation,
  in the matter-dominated and radiation-dominated universe, or more
  generally in the universe with the equation of state $p=w\rho$. We
  have found that the domain wall evolution crucially depends on the
  time-dependent parameter $C(t)=1/(H(t)\delta_0)^2$, where $H(t)$ is
  the Hubble parameter and $\delta_0$ is the thickness of the wall in flat
  space-time. For $C(t)>2$ the physical thickness of the wall,
  $a(t)\delta(t)$, tends with time to $\delta_0$, which
  is microscopically small.  Otherwise, when $C(t) \leq 2$, the wall
  steadily expands and can grow up to a cosmologically large size.  
\end{abstract}

\maketitle


\section{Introduction}
\label{sec:Intro}

Creation of the baryon asymmetry of the universe and possible
existence of the cosmological antimatter crucially depends upon the
version of $C$ and $CP$ violation realized in the early universe. In
this connection spontaneous $CP$ violation suggested in
ref.~\cite{lee-CP} is of particular interest. This beautiful model,
however, suffers from the domain wall problem~\cite{zko} -- the energy
density of the walls between domains with different sign of $CP$
violation is so large that they would either overclose the universe or
destroy the observed (near-)isotropy of the microwave background
radiation. To avoid this problem the mechanism of the wall destruction
was proposed, see e.g.~\cite{dgrt} and references therein. If this
mechanism is operative, it should get rid of remnants of the walls not
earlier then the baryogenesis was accomplished and the universe
acquired the necessary baryon asymmetry. Double-well shape of the
potential would already vanish to this period and the $CP$-violating
scalar field $\varphi$, which formed the domain wall, would evolve down to zero. 
Baryogenesis in
this scenario took place when $\varphi$ rolled down to zero, but still
before it reached the mechanical equilibrium point at $\varphi = 0$. 
Different related mechanisms of $C$ and $CP$ violation in cosmology
are reviewed in ref.~\cite{AD-CP-cosm}.

Another problem of the baryogenesis based on spontaneous $CP$
violation is the thickness of the wall.  In flat space-time the thickness of
the wall is microscopically small and if the walls with such or
similar thicknesses were created in the cosmological situation, the
matter-antimatter domains would be in close contact with each other.
It would lead to very large annihilation rate and to unacceptably high
background of the annihilation products in the universe. However, the
cosmological expansion may lead to much larger separation of the
domains eliminating or smoothing down this problem.

The evolution of the domain wall thickness in de Sitter universe, 
which is an approximation to the inflationary universe, was studied in the
papers~\cite{BV, DGR}. It was shown there that for sufficiently small
value of the parameter
\begin{equation}
  C(t) =1/(H(t)\delta_0)^2,
  \label{C1}
\end{equation}
where $H(t)$ is the Hubble parameter and $\delta_0$ is the thickness of
the wall in flat space-time, the wall thickness
would exponentially rise and the matter-antimatter domains may be
safely separated. 

In this work we studied the evolution of thick domain walls in the different models 
of inflation, as well as for arbitrary
cosmological expansion regimes with the matter
satisfying the equation of state $p = w \rho$ with constant
parameter $w$. We have shown that there exists some range of the
inflation parameters leading to a large domain separation prior to
successful baryogenesis.


\section{Evolution of thick domain walls in de Sitter universe}
\label{sec:deSitter}

In this section we briefly remind how the thick domain walls evolve in
de Sitter universe~\cite{BV, DGR}.

We consider a model of real scalar field $\varphi$ with a simple
double-well potential. The Lagrangian of such model is the following,
\begin{equation}
\mathcal{L} = \frac{1}{2}g^{\mu\nu}
\partial_\mu\varphi\, \partial_\nu\varphi -
\frac{\lambda}{2}\left(\varphi^2-\eta^2\right)^2.
\label{eq:lagrangian}
\end{equation}

The equation of motion of field $\varphi$ can be easily obtained
from~(\ref{eq:lagrangian}) and looks as
\begin{equation} \label{eq_of_mot}
\frac{1}{\sqrt{-g}} \partial_\mu \left(\sqrt{-g} g^{\mu\nu}
\partial_\nu\varphi \right) = -2\lambda\varphi\left(\varphi^2-\eta^2\right).
\end{equation}

\begin{figure}[t]
\center
\includegraphics[width=4.5in]{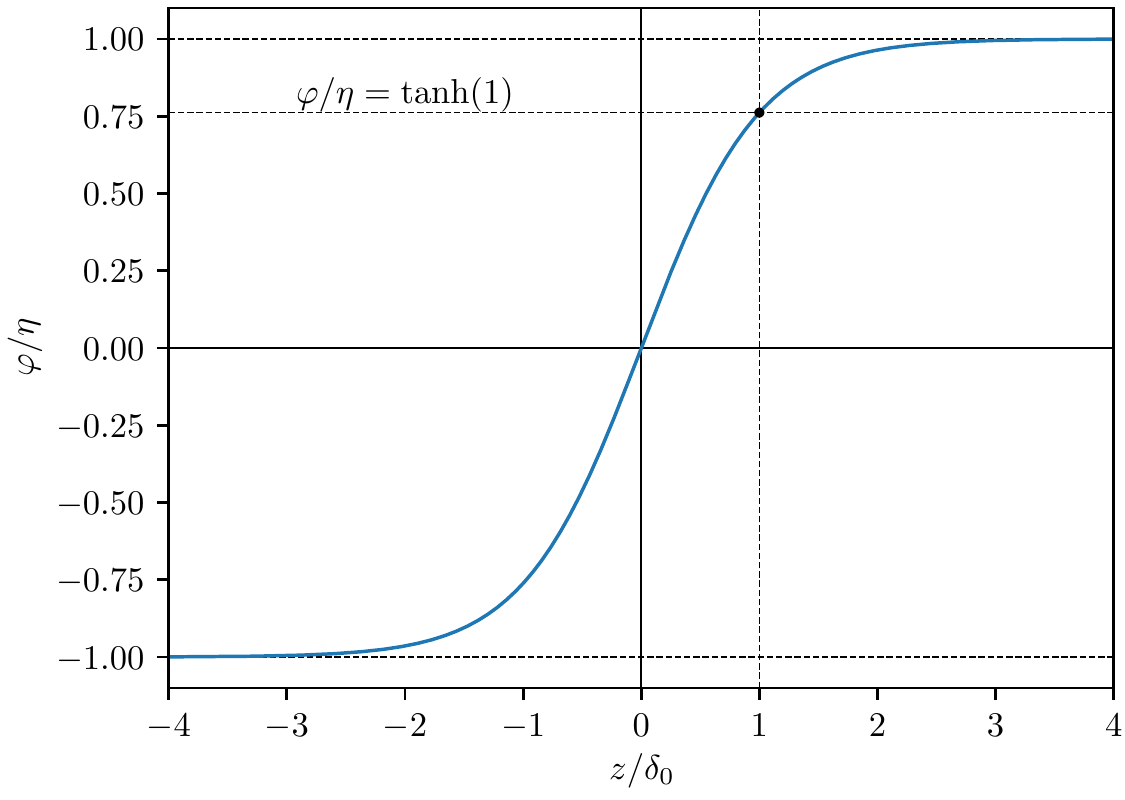}
\caption{\label{fig:tanh} 
Domain wall in flat space-time.}
\end{figure}

In flat space-time with the metric,
$ds^2 = dt^2-\left(dx^2+dy^2+dz^2\right)$, and in static
one-dimensional case, $\varphi=\varphi(z)$, the equation
(\ref{eq_of_mot}) has the form,
\begin{equation}
\frac{d^2\varphi}{dz^2}=2\lambda\varphi\left(\varphi^2-\eta^2\right).
\end{equation}
This equation has a kink-type solution, describing a static infinite
flat domain wall in $xy$-plane.  Without loss of generality we can
assume that the wall is situated at $z=0$,
\begin{equation}
  \label{flat}
  \varphi(z)=\eta\,\tanh{\frac{z}{\delta_0}},
\end{equation}
where $\delta_0=1/(\sqrt{\lambda}\eta)$ is the thickness of the wall in
flat space-time (see Fig.~\ref{fig:tanh}).
Since $\sqrt{\lambda}\eta$ is essentially the mass of the Higgs-like boson, 
$\delta_0$ is microscopically small. 
Otherwise, if $\delta_0$ would be cosmologically large, this boson would have a tiny mass and thus it would generate long range forces which are strongly restricted by experiment.

The evolution of thick domain walls in spatially flat section of the
de Sitter universe with the metric
$ds^2 = dt^2-e^{2Ht}\left(dx^2+dy^2+dz^2\right)$, with a constant
Hubble parameter, $H>0$, was considered previously in
refs.~\cite{BV,DGR}. A remarkable feature of this problem is that all
parameters can be combined into a single positive constant,
$C=1/(H\delta_0)^2=\lambda\eta^2/H^2>0$.  Therefore, it is quite
natural that the evolution of domain walls is determined only by the
value of $C$.  In the case of very thin domain walls, whose thickness
is much smaller than the de Sitter horizon, $\delta \ll H^{-1}$, i.e.
$C \gg 1$, the solution $\varphi(z)$ is well approximated by the
flat-spacetime solution~(\ref{flat}).  However, as the flat-spacetime
thickness parameter $\delta_0$ increases, a deviation of the solution
from the flat-spacetime solution increases as well.  For sufficiently
large value of parameter $C$, $C > 2$, the initial kink configuration
in a de Sitter background tends to the stationary solution which
depends only on physical distance $l=a(t)z$, i.e.
$\varphi =\eta f\left(Hl\right) = \eta f(e^{Ht} Hz)$. The thickness of the
stationary wall rises with decreasing value of~$C$.

Above the critical value, $\delta_0 \geq H^{-1}/\sqrt{2}$, i.e.
$C \leq 2$, there are no stationary solutions at all.  But if one
allows for an arbitrary dependence of the solution on $z$ and $t$, the
solution exists for any $C$, and the case of $C\leq 2$ leads to the
expanding kink with rising thickness. In other words, the thickness of the
wall infinitely grows with time. For $C\lesssim 0.1$ the rise is close
to the exponential one, therefore transition regions between domains
might be cosmologically large.


\section{Evolution of thick domain walls in inflationary universe}
\label{sec:inflation}

\subsection{$\Phi^2$-inflation}

\begin{figure}[t]
  \centering
  \includegraphics[width=4.5in]{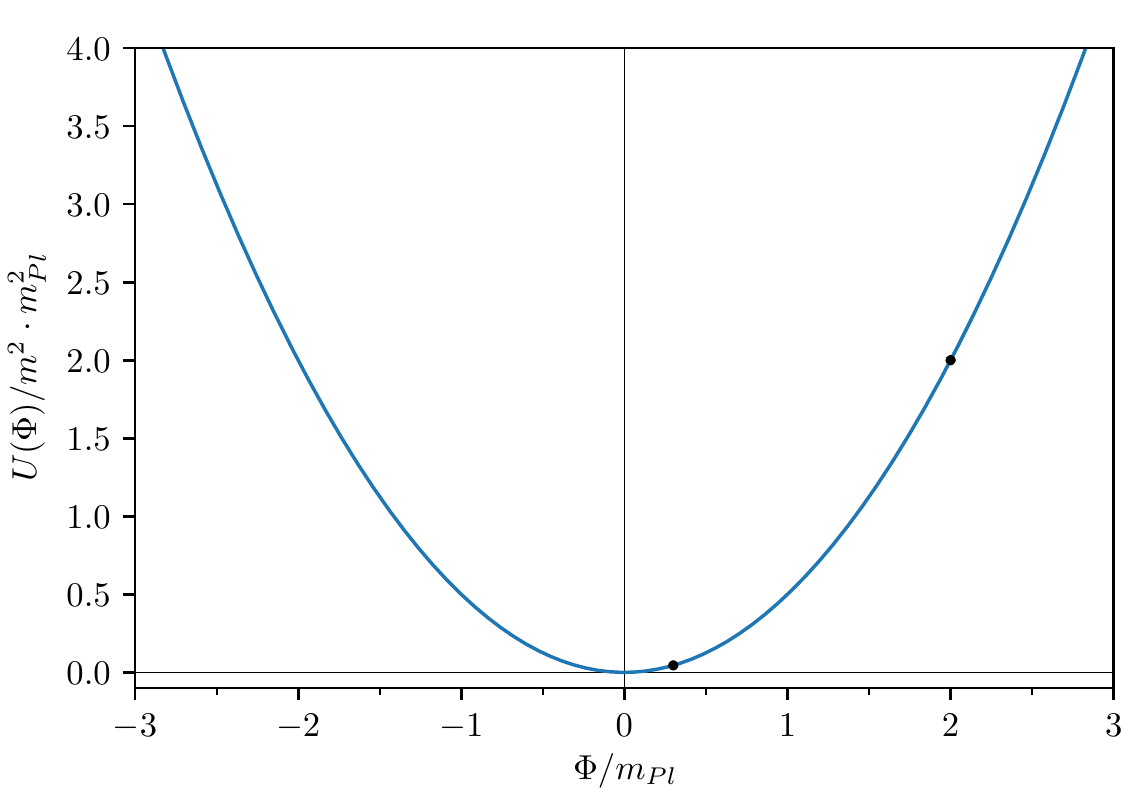}
  \caption{Inflaton potential $U(\Phi)$ in the $\Phi^{2}$-model. Black
    dots correspond to $\Phi_i=2.0\, m_{Pl}$ and
    $\Phi_e=0.3\, m_{Pl}$.}
  \label{fig:Phi2_U}
\end{figure}

In this section we consider a simple model of inflation with quadratic
inflaton potential~(see Fig.~\ref{fig:Phi2_U})
\begin{equation}
  U(\Phi)=\frac{m^2\Phi^2}{2},
\end{equation}
(the inflaton field $\Phi$ should not be confused with the field
$\varphi$ which forms the domain wall).  The model of inflation with a
quadratic potential is strongly constrained in view of the recent
observational data (for a review see e.g.~\cite{PDG:2016}), but for
our conclusion only the very fact of exponential expansion is
important, regardless of the specific mechanism of inflation.

We assume that potential energy $U$ of the inflaton gives the main
contribution to the cosmological energy density $\rho$.  Therefore,
the Hubble parameter and the scale factor are completely determined by
the inflaton field.

We consider the evolution of domain wall during the slow-roll regime
of inflation, when the inflaton field $\Phi$ slowly rolls down from
some initial value $\Phi_i$ to the minimum $\Phi=0$ of the potential
$U(\Phi)$.  Slow-roll regime ends when at least one of the slow-roll
parameters $\epsilon(\Phi)$ or $\eta(\Phi)$ becomes of the order of 1.
In the model under consideration the slow-roll parameters coincide,

\begin{align}
  \epsilon(\Phi)&=\frac{m_{Pl}^2}{16\pi}\left(\frac{U'(\Phi)}{U(\Phi)}\right)^2=\frac{1}{4\pi}\frac{m_{Pl}^2}{\Phi^2},\\
  \eta(\Phi)&=\frac{m_{Pl}^2}{8\pi}\frac{U''(\Phi)}{U(\Phi)}=\frac{1}{4\pi}\frac{m_{Pl}^2}{\Phi^2}=\epsilon(\Phi),
\end{align}
where $m_{Pl}\approx 1.2\cdot 10^{19}$ GeV is the Planck mass.
Therefore, the slow-roll regime of inflation ends when
$\Phi \simeq m_{Pl}/\sqrt{4\pi}\approx 0.3\, m_{Pl}$.
It should be noted that in the models of inflation that we consider in this paper
the end of slow-roll regime means the end of inflation itself.

In contrast to de Sitter universe (see Section~\ref{sec:deSitter}) the
Hubble parameter now depends on time,
\begin{equation}
  H(t)=\sqrt{\frac{8\pi\rho(t)}{3m_{Pl}^2}} \approx \sqrt{\frac{8\pi}{3m_{Pl}^2}\frac{m^2 \Phi^2(t)}{2}} = 
  \sqrt{\frac{4\pi}{3}}\frac{m}{m_{Pl}}\Phi(t) \neq const,
\end{equation}
so at the end of the slow-roll regime, $H \simeq m/\sqrt{3}$.

The equation of motion of the inflaton in the slow-roll regime is the
following,
\begin{equation}
  \dot{\Phi}(t) \approx -\frac{m^2 \Phi(t)}{3H(t)} \approx -\frac{m_{Pl} m}{\sqrt{12\pi}}.
\end{equation}
This equation is easily integrated,
\begin{equation}
  \Phi (t) =  \Phi_i - \frac{m_{Pl} m}{\sqrt{12\pi}}t,
\end{equation}
where $\Phi_i$ is the initial value of the inflaton field.

The Hubble parameter and the scale factor can also be easily found,
\begin{align}
  H(t) &= \sqrt{\frac{4\pi}{3}}\frac{m}{m_{Pl}}\Phi_i - \frac{1}{3}m^2 t,
         \label{eq:inf_H}\\
  a(t) &= a_0\cdot \exp{\left( \sqrt{\frac{4\pi}{3}}\frac{m}{m_{Pl}}\Phi_i t - \frac{1}{6}m^2 t^2 \right)}.
\end{align}
These formulas are valid only till the end of the slow-roll regime,
\begin{equation}
  \label{te}
  t \lesssim t_e =
  \left(\sqrt{12\pi}\frac{\Phi_i}{m_{Pl}} - \sqrt{3}\right)m^{-1}.
\end{equation}

The equation of motion~(\ref{eq_of_mot}) in the case when the field
$\varphi$ is a function of two independent variables, $z$ and $t$, is
written as
\begin{equation}
  \label{eq-mot-f}
  \frac{\partial^2 f}{\partial
    t^2}+3H(t)\frac{\partial f}{\partial
    t}-\frac{1}{a^2(t)}\frac{\partial^2 f}{\partial z^2}
  =\frac{2}{\delta_0^2} f \left(1-f^2\right),
\end{equation}
where $f(z,t)=\varphi(z,t)/\eta$.

Since $H(t)$ has the form~(\ref{eq:inf_H}), it is convenient to use
$1/m$ units in equation of motion (\ref{eq-mot-f}):
\begin{align}
  &H(t)=\frac{m}{3}(mt_e-mt)+\frac{m}{\sqrt{3}},\\
  &a(t)=a_e \cdot \exp{\left(
    -\frac{(mt_e-mt)^2}{6}-\frac{(mt_e-mt)}{\sqrt{3}}\right)} \hspace{3mm} \text{with} \hspace{3mm} 
    a_e=a_0\cdot e^{m^2 t_e^2/6+mt_e/\sqrt{3}},\label{eq:sf_inf}\\
  &\label{eq:inf_eom}
    \frac{\partial^2 f}{\partial
    \left(t\cdot m\right)^2}+\left((mt_e-mt)+\sqrt{3}\right)
    \frac{\partial f}{\partial
    \left(t\cdot m\right)}-\frac{1}{a^2(t)}\frac{\partial^2
    f}{\partial \left(z\cdot m\right)^2}
    =\frac{2}{\left(m\cdot\delta_0\right)^2} f \left(1-f^2\right).
\end{align}

The boundary conditions for the kink-type solution should be
\begin{equation}
  \label{bound_cond1}
  f(0,t)=0, \hspace{5mm}
  f(\pm\infty,t)=\pm1,
\end{equation}
and we choose the initial configuration as the domain wall with
physical thickness $\delta_0$ and zero time derivative:
\begin{equation}
  \label{in_cond1}
  f(z,t_i)=\tanh{\frac{z\cdot a\left(t_{i}\right)}{\delta_0}},
  \hspace{5mm}
  \frac{\partial f(z,t)}{\partial t}\biggl|_{t=t_i}=0.
\end{equation}
In numerical calculations we use the following values: 
$\Phi_i=2\,m_{Pl}$, $t_i=0$, and $a_0=1$. 

\begin{figure}[ht]
  \centering
  \begin{subfigure}[b]{0.493\textwidth}
    \centering
    \includegraphics[width=3.0in]{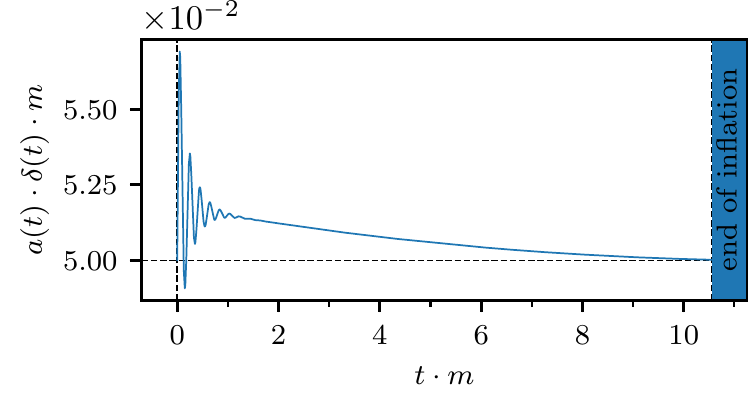}
    \caption{$\delta_0=0.05\cdot m^{-1}$}
    \label{fig:md=0.050}
  \end{subfigure}
  \begin{subfigure}[b]{0.493\textwidth}
    \centering
    \includegraphics[width=3.0in]{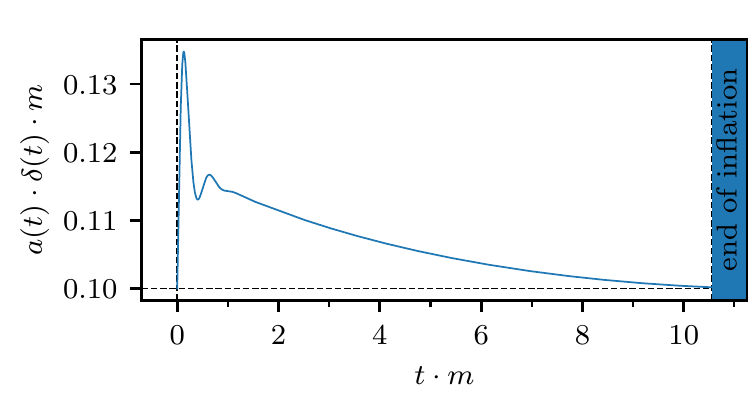}
    \caption{$\delta_0=0.1\cdot m^{-1}$}
    \label{fig:md=0.100}
  \end{subfigure}\\
  \begin{subfigure}[b]{0.493\textwidth}
    \centering
    \includegraphics[width=3.0in]{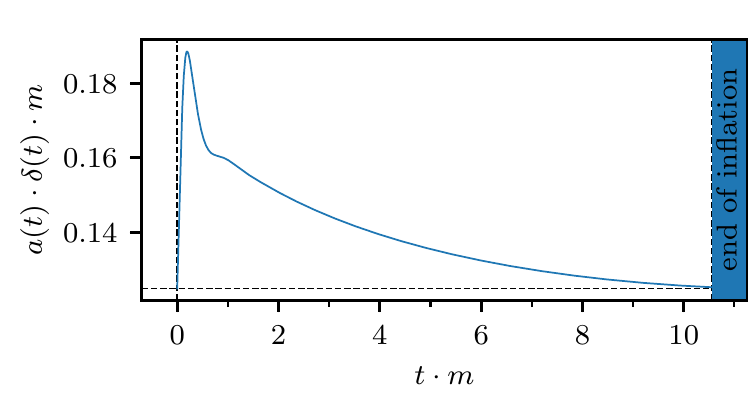}
    \caption{$\delta_0=0.125\cdot m^{-1}$}
    \label{fig:md=0.125}
  \end{subfigure}
  \begin{subfigure}[b]{0.493\textwidth}
    \centering
    \includegraphics[width=3.0in]{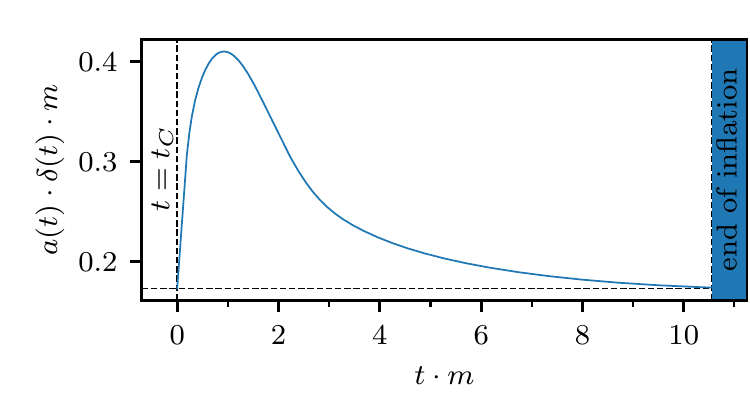}
    \caption{$\delta_0\approx 0.173\cdot m^{-1}$ $(t_C=0)$}
    \label{fig:md=0.173}
  \end{subfigure}\\
  \begin{subfigure}[b]{0.493\textwidth}
    \centering
    \includegraphics[width=3.0in]{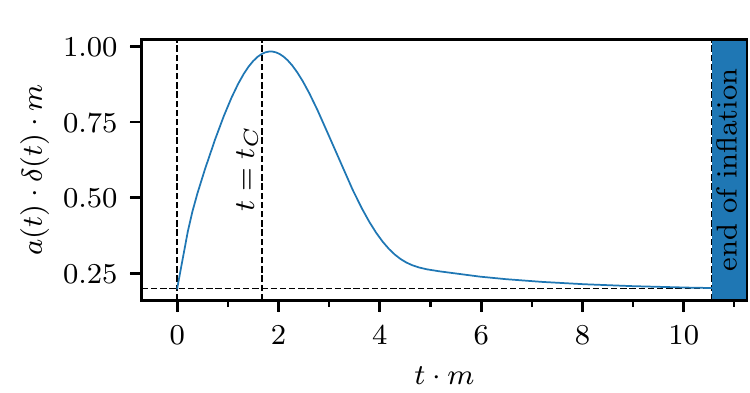}
    \caption{$\delta_0=0.2\cdot m^{-1}$}
    \label{fig:md=0.200}
  \end{subfigure}
  \begin{subfigure}[b]{0.493\textwidth}
    \centering
    \includegraphics[width=3.0in]{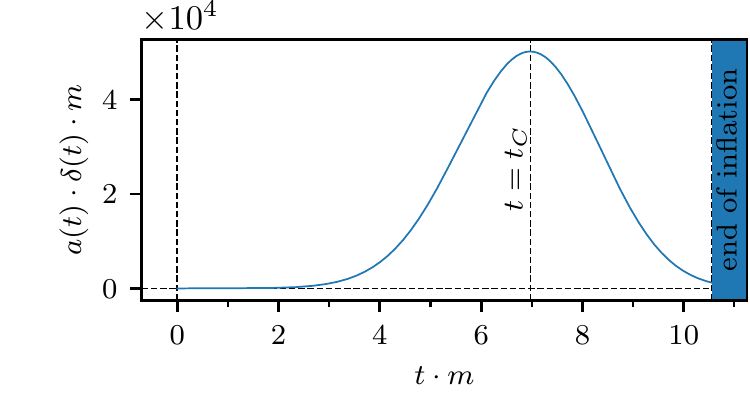}
    \caption{$\delta_0=0.4\cdot m^{-1}$}
    \label{fig:md=0.400}
  \end{subfigure}\\
  \begin{subfigure}[b]{0.493\textwidth}
    \centering
    \includegraphics[width=3.0in]{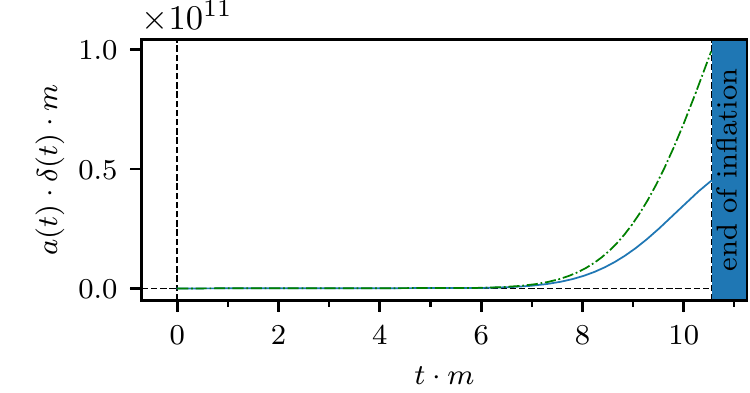}
    \caption{$\delta_0=2.0\cdot m^{-1}$}
    \label{fig:md=2.000}
  \end{subfigure}
  \begin{subfigure}[b]{0.493\textwidth}
    \centering
    \includegraphics[width=3.0in]{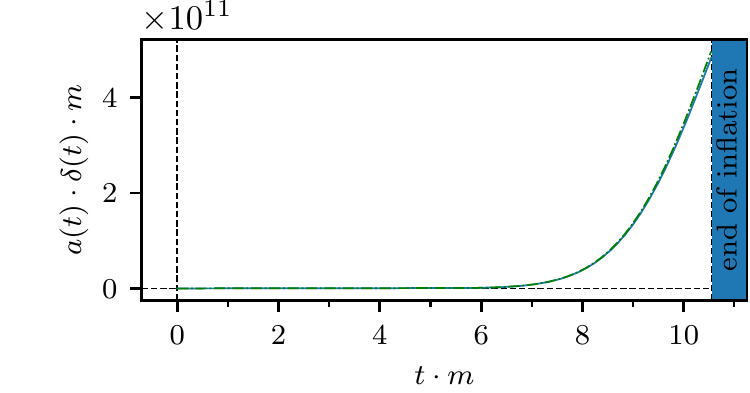}
    \caption{$\delta_0=10\cdot m^{-1}$}
    \label{fig:md=10.000}
  \end{subfigure}
  \caption{Time dependence of physical thickness of the wall,
    $a(t)\cdot\delta(t)\cdot m$, for different values of the initial
    wall thickness, $\delta_0$, in the $\Phi^2$-model.  
    Dashed horizontal line corresponds to
    $\delta_0$. The time at which the inflation ends is shown by
    vertical dashed line with the corresponding label.  In
    Figs.~\ref{fig:md=0.173}-\ref{fig:md=0.400} there is a vertical
    dashed line which corresponds to $t=t_{C}$. In
    Figs.~\ref{fig:md=2.000} and \ref{fig:md=10.000} there is a green
    (dash-dotted) line which corresponds to
    $a(t)\cdot\delta_0\cdot m$.  In Fig.~\ref{fig:md=10.000} this line
    coincides with $a(t)\cdot\delta(t)\cdot m$, which means that
    $\delta\left(t\right)\approx \delta_0$ till the end of inflation
    with very good accuracy.}
  \label{fig:inflation_all}
\end{figure}

Time dependence of physical thickness of the wall, $a(t)\delta(t)$, for
different values of the initial wall thickness, $\delta_0$, is shown in
Fig.~\ref{fig:inflation_all}. Here $\delta(t)$ is the
coordinate thickness of the wall. It is defined as the value of
the coordinate $z$ at the position where $f(z,t)=\tanh{1}\approx 0.76$.

It turns out that the domain wall evolution is basically determined by the parameter
\begin{equation}
C(t)=\frac{1}{(H(t)\delta_0)^2}.
\end{equation}
Since $H(t)$ is decreasing, $C(t)$ increases with time.
As is known from Section~\ref{sec:deSitter}, when
$C<2$ the thickness of the wall increases rapidly while for $C>2$ the wall
thickness tends to its stationary value which for $C\gg2$ coincides with
$\delta_0$.

Time $t_C$ at which $C(t_C)=2$ in the model under consideration is determined from the
relation
\begin{equation}
  m t_C=\sqrt{12\pi}\frac{\Phi_i}{m_{Pl}}-\frac{3}{\sqrt{2}m\delta_0}.
  \label{eq:inf_tC}
\end{equation}
Parameter $C(t)$ can be equal 2 only if $t_C \geq t_i=0$, i.e.
if $m \delta_0 \geq \sqrt{3}m_{Pl}/(\sqrt{8\pi}\Phi_i)\approx 0.173$
for our choice of $\Phi_i$.
From (\ref{te}) and (\ref{eq:inf_tC}) it follows that $t_C < t_e $ if 
$m \delta_0 < \sqrt{3/2}\approx 1.225$.

When $\delta_0$ is so small that $C(t)>2$ during all the time of inflation, 
i.e. $m \delta_0 < \sqrt{3}m_{Pl}/(\sqrt{8\pi}\Phi_i)$, then the domain wall thickness, 
$a(t)\delta(t)$, after some
damped oscillations tends to constant value, $\delta_0$, which is
microscopically small. 

For larger $\delta_0$, such that $\sqrt{3}m_{Pl}/(\sqrt{8\pi}\Phi_i) \leq m \delta_0 \leq \sqrt{3/2}$,
initially one has $C(t)\leq 2$, but at the end of the slow-roll regime $C(t)\geq 2$. 
Therefore, the domain wall thickness grows initially, reaches the maximum and then diminishes.

Finally, if $\delta_0$ is quite large, $m \delta_0 > \sqrt{3/2}$,
then $C(t)<2$ during all the time of inflation. 
In such case domain wall thickness could
grow to cosmologically large size by the end of inflation. 
However, for domain walls, which were thick
initially, $m\delta_0\gg 1$, coordinate thickness almost does not change,
$\delta(t)\approx \delta_0$, and domain wall expansion is entirely due
to growth of scale factor, $a(t)$
(see Fig.~\ref{fig:md=10.000}).


\subsection{``Hilltop''-inflation}

\begin{figure}[t]
  \centering
  \includegraphics[width=4.5in]{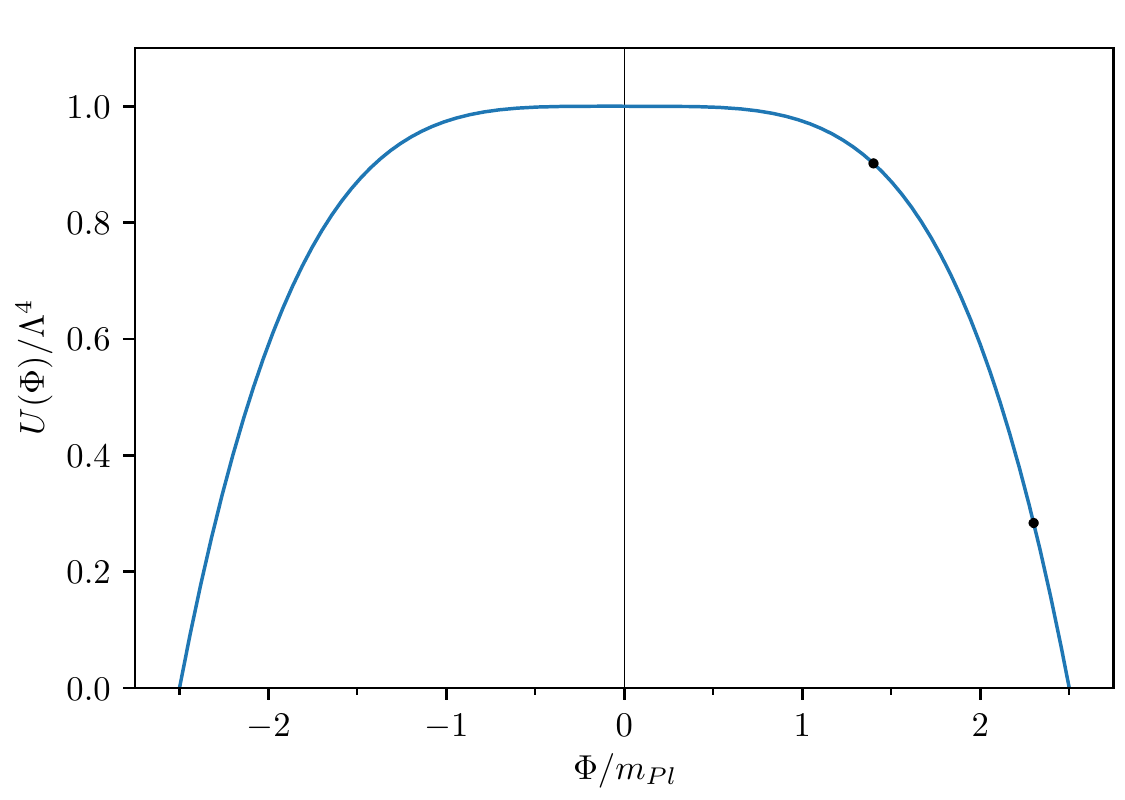}
  \caption{Inflaton potential $U(\Phi)$ in the ``hilltop''
    model. Black dots correspond to $\Phi_i=1.4\, m_{Pl}$ and
    $\Phi_e=2.3\, m_{Pl}$.}
  \label{fig:hilltop_U}
\end{figure}

Since the $\Phi^2$-inflation is in tension with recent observational
data let us consider below the other inflation models which are in agreement with observations.
One of such models is the so-called ``hilltop'' model~\cite{Boubekeur:2005}.  
The inflaton potential for quite small values of $\Phi$ looks as follows (see Fig.~\ref{fig:hilltop_U})
\begin{equation}
U(\Phi)= \Lambda^4\left(1-\frac{\Phi^4}{\mu^4}\right).
\end{equation}
We take $\mu=2.5\, m_{Pl}$ to be consistent with the Planck data~\cite{Planck:2015}.

The slow-roll parameters are
\begin{align}
\epsilon(\Phi)&=\frac{m_{Pl}^2}{16\pi}\left(\frac{U'(\Phi)}{U(\Phi)}\right)^2=\frac{m_{Pl}^2}{\pi}\frac{\Phi^6}{(\mu^4-\Phi^4)^2},\\
\eta(\Phi)&=\frac{m_{Pl}^2}{8\pi}\frac{U''(\Phi)}{U(\Phi)}=-\frac{3 m_{Pl}^2}{2 \pi}\frac{\Phi^2}{\mu^4-\Phi^4}.
\end{align}

It is seen that $\epsilon(\Phi), |\eta(\Phi)| \ll 1$ for
$\Phi \ll \mu \sim m_{Pl}$. The slow-regime of inflation ends when
$\epsilon(\Phi) \sim 1$ or $|\eta(\Phi)| \sim 1$, so in this model we
take $\Phi_e=2.3\, m_{Pl}$.

The Hubble parameter is 
\begin{equation}
H(t)\simeq \sqrt{\frac{8\pi}{3m_{Pl}^2}U(\Phi(t))},
\end{equation}
and correspondingly the scale factor is
\begin{equation}
a(t)=a_0\cdot {\rm exp}\left(\int_0^t{H(t')dt'}\right),
\end{equation}
in the calculations we use $a_0=1$, therefore $a(0)=1$.

We take $\Lambda=10^{-3}\, m_{Pl}$ 
in agreement with measured value of the amplitude of the scalar density perturbations,
$\Delta_R\simeq 5\cdot 10^{-5}$,
\begin{equation}
  \Delta_R=\frac{3H^3}{2\pi |U'(\Phi)|}=
  \sqrt{\frac{8\pi}{3}}\frac{\Lambda^2\mu^4}{\Phi^3 m_{Pl}^3}\left(1-\frac{\Phi^4}{\mu^4}\right)^\frac{3}{2},
\end{equation}
here the value of $\Phi$ should be taken at the moment of 50--60
e-foldings before the end of inflation.

The number of e-foldings till the end of inflation is
$N_e(\Phi)\simeq\pi\mu^4/(m_{Pl}^2\Phi^2)$ for
$\Phi \ll \mu \sim m_{Pl}$. Initial value $\Phi_0=0$ leads to
divergence, so in such models one takes $\Phi_0=H(0)\ll m_{Pl}$. In
such a case $N_e(\Phi_0)\simeq 10^{13}$, so the condition $N>60$ of
successful inflation is well fulfilled.
We choose the time when the domain wall was formed not at the
very beginning of inflation but rather near the end of inflation,
$\Phi_i=\Phi(0)=1.4\, m_{Pl}$ (see Fig.~\ref{fig:hilltop_U}).

Solution $\Phi(t)$ of the equation of motion,
$3H(t)\dot{\Phi}(t)\approx -U'(\Phi)$, during the slow-roll regime of
inflation can be expressed as implicit function,
\begin{equation}
  \arctan{\sqrt{\frac{\mu^4}{\Phi^4(t)}-1}}-\sqrt{\frac{\mu^4}{\Phi^4(t)}-1}=\sqrt{\frac{8}{3\pi}}\frac{\Lambda^2}{\mu^2}m_{Pl} t + \arctan{\sqrt{\frac{\mu^4}{\Phi_i^4}-1}}-\sqrt{\frac{\mu^4}{\Phi_i^4}-1}.
\end{equation}
From this equation we can find the duration of the inflation for our
choice of parameters: $t_{e}\approx 11.6\cdot m_{Pl}/\Lambda^{2}$.

Since $U\left(\Phi\right)\propto \Lambda^{4}$, the Hubble parameter
$H\propto \Lambda^{2}/m_{Pl}$. Therefore, $m_{Pl}/\Lambda^{2}$ is the
natural choice of units for $t$ and $\delta\left(t\right)$ in this
model, like $m^{-1}$ units in $\Phi^{2}$-model, see~(\ref{eq:inf_eom}).

Now we can find numerically $\Phi(t)$, $H(t)$, and $a(t)$ at given
$t$, and therefore solve the equation describing the evolution of the
field $\varphi$ (\ref{eq-mot-f}).  The boundary conditions as usually
are (\ref{bound_cond1}) and we choose the initial configuration as the
domain wall with physical thickness $\delta_0$ and zero time
derivative (\ref{in_cond1}).

As it was in the $\Phi^{2}$-model, we expect that the evolution of the
wall is defined by $C\left(t\right)$. 
The values of $\delta_0$ for which $C(t)$ can be equal to 2 during the wall evolution
can be found from the inequality $0 \leq t_C \leq t_e $, and lie between the following values:
\begin{align}
  &t_{C}=0\text{ for
    }\delta_{0}=\frac{1}{H\left(0\right)\sqrt{2}}=\sqrt{\frac{3}{16\pi\left(1-\frac{\Phi_{i}^{4}}{\mu^{4}}\right)}}\cdot\frac{m_{Pl}}{\Lambda^{2}}
    \approx 0.2573\cdot\frac{m_{Pl}}{\Lambda^{2}},\\
  &t_{C}=t_{e}\text{ for
    }\delta_{0}=\frac{1}{H\left(t_{e}\right)\sqrt{2}}=\sqrt{\frac{3}{16\pi\left(1-\frac{\Phi_{e}^{4}}{\mu^{4}}\right)}}\cdot\frac{m_{Pl}}{\Lambda^{2}}
    \approx 0.4587\cdot\frac{m_{Pl}}{\Lambda^{2}}.
\end{align}

\begin{figure}[t]
  \centering
  \begin{subfigure}[b]{0.493\textwidth}
    \centering
    \includegraphics[width=3.0in]{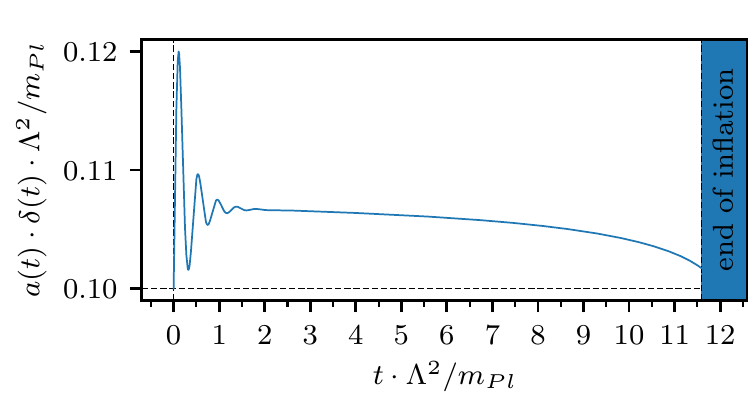}
    \caption{$\delta_0=0.1\cdot m_{Pl}/\Lambda^{2}$}
    \label{fig:hilltop=0.1000}
  \end{subfigure}
  \begin{subfigure}[b]{0.493\textwidth}
    \centering
    \includegraphics[width=3.0in]{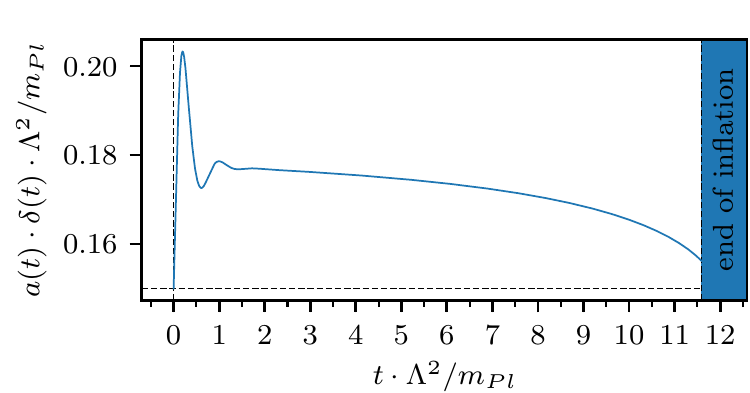}
    \caption{$\delta_0=0.15\cdot m_{Pl}/\Lambda^{2}$}
    \label{fig:hilltop=0.1500}
  \end{subfigure}\\
  \begin{subfigure}[b]{0.493\textwidth}
    \centering
    \includegraphics[width=3.0in]{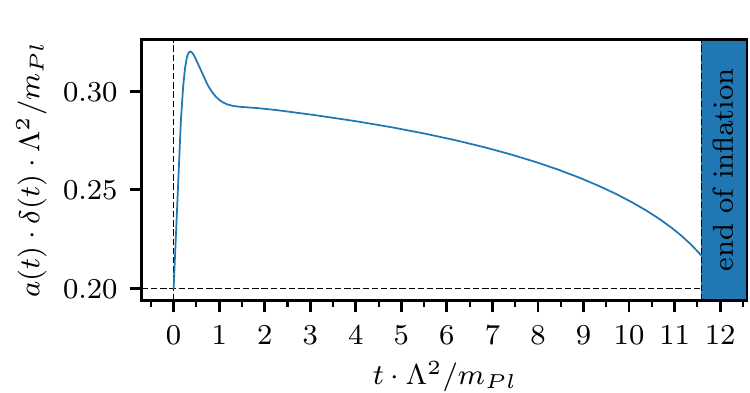}
    \caption{$\delta_0=0.2\cdot m_{Pl}/\Lambda^{2}$}
    \label{fig:hilltop=0.2000}
  \end{subfigure}
  \begin{subfigure}[b]{0.493\textwidth}
    \centering
    \includegraphics[width=3.0in]{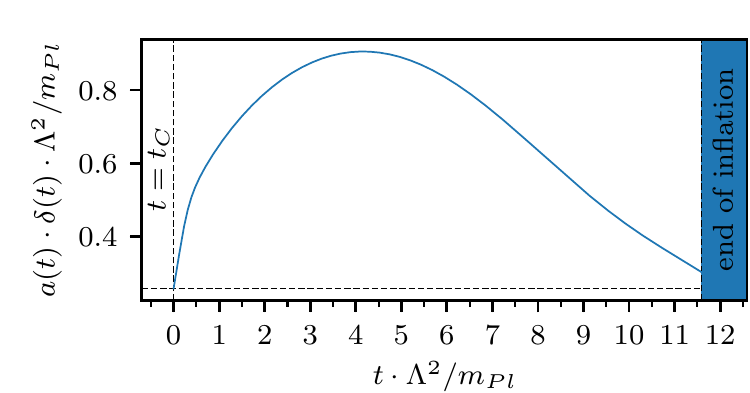}
    \caption{$\delta_0\approx0.2573\cdot m_{Pl}/\Lambda^{2}$ $(t_C=0)$}
    \label{fig:hilltop=0.2573}
  \end{subfigure}\\
  \begin{subfigure}[b]{0.493\textwidth}
    \centering
    \includegraphics[width=3.0in]{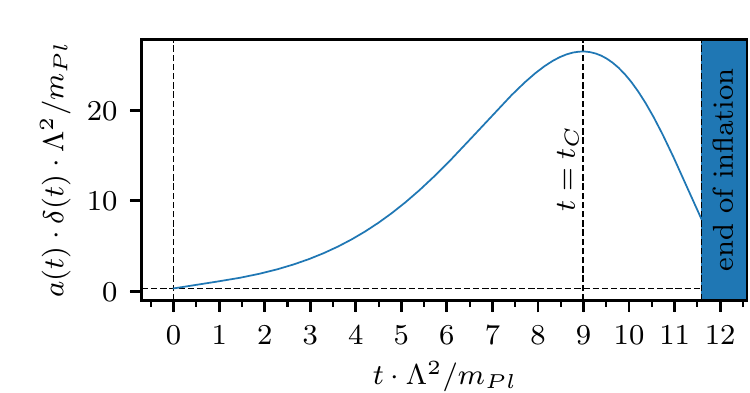}
    \caption{$\delta_0=0.3\cdot m_{Pl}/\Lambda^{2}$}
    \label{fig:hilltop=0.3000}
  \end{subfigure}
  \begin{subfigure}[b]{0.493\textwidth}
    \centering
    \includegraphics[width=3.0in]{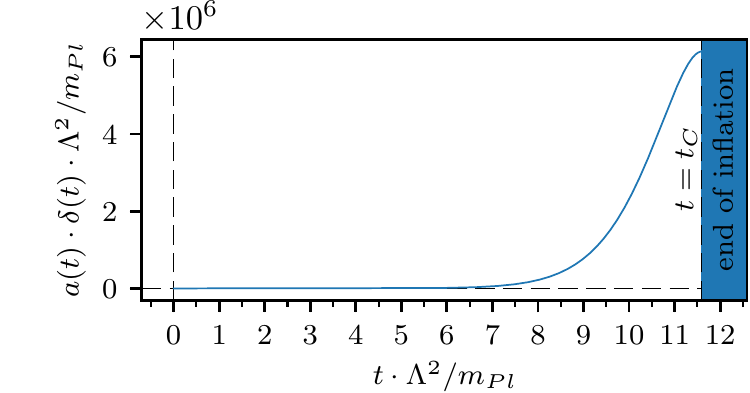}
    \caption{$\delta_0\approx0.4587\cdot m_{Pl}/\Lambda^{2}$ $(t_C=t_e)$}
    \label{fig:hilltop=0.4587}
  \end{subfigure}\\
  \begin{subfigure}[b]{0.493\textwidth}
    \centering
    \includegraphics[width=3.0in]{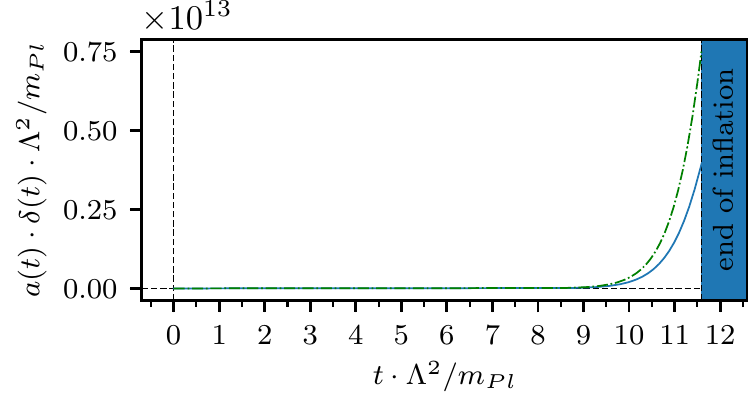}
    \caption{$\delta_0=2.0\cdot m_{Pl}/\Lambda^{2}$}
    \label{fig:hilltop=2.0000}
  \end{subfigure}
  \begin{subfigure}[b]{0.493\textwidth}
    \centering
    \includegraphics[width=3.0in]{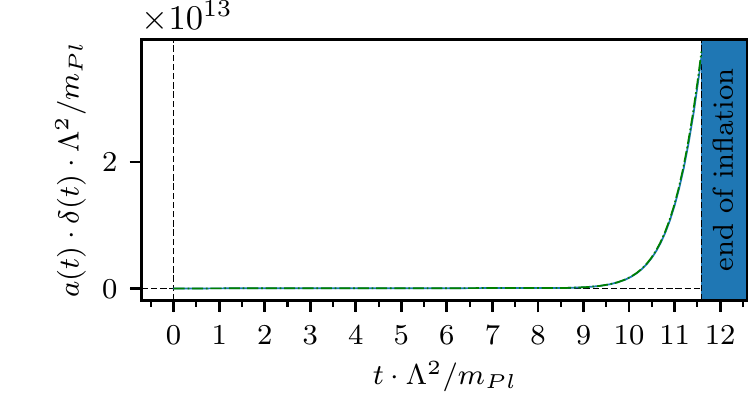}
    \caption{$\delta_0=10.0\cdot m_{Pl}/\Lambda^{2}$}
    \label{fig:hilltop=10.0000}
  \end{subfigure}
  
  \caption{Time dependence of physical thickness of the wall,
    $a(t)\cdot\delta(t)$, for different values of the initial wall
    thickness, $\delta_0$, in the ``hilltop'' model.  Here $t$,
    $a(t)\delta(t)$ and $\delta_0$ are shown in
    $m_{Pl}/\Lambda^{2}=(10^{-6}m_{Pl})^{-1}$ units. Dashed horizontal
    line corresponds to $\delta_0$. The time at which the inflation
    ends is shown by vertical dashed line with the corresponding
    label. In Figs.~\ref{fig:hilltop=0.2573}--\ref{fig:hilltop=0.4587} there is a vertical dashed line which
    corresponds to $t=t_C$. In Figs.~\ref{fig:hilltop=2.0000} and
    \ref{fig:hilltop=10.0000} there is a green (dash-dotted) line
    which corresponds to $a(t)\cdot\delta_0\cdot m$.}
  \label{fig:hilltop_all}
\end{figure}

The results of numerical calculation for the time dependence of the
physical thickness of the wall, $a(t)\delta(t)$, for different values of
parameter $\delta_0$ are presented in Fig.~\ref{fig:hilltop_all}.

The results are similar to what we obtained in $\Phi^{2}$-model. Very
thin domain walls do not expand significantly, oscillating near
$\delta_{0}$, see
Figs.~\ref{fig:hilltop=0.1000}--\ref{fig:hilltop=0.2000}. For a wall
of medium size, such that $t_C$ is after the beginning of inflation but before the end, there are periods when the wall thickness increases and
decreases very fast,
see~Figs.~\ref{fig:hilltop=0.2573},~\ref{fig:hilltop=0.3000}. Finally,
a very thick wall, such that $t_{C}$ is after the end of inflation,
grows with the universe, see~Fig.~\ref{fig:hilltop=10.0000}.

\subsection{$R^2$-inflation}

One more inflation model consistent with observational data is
$R^2$-inflation proposed and developed by Starobinsky in the late
1970s~\cite{Starobinsky:1980, Starobinsky:1981} (for review see
e.g.~\cite{DeFelice:2010}).  The modified theory of gravity with the
action
\begin{equation}
  S = -\frac{m_{Pl}^2}{16\pi}\int{d^4 x \sqrt{-g} \left(R-\frac{R^2}{6M^2}\right)}
\end{equation}
leads to the inflationary expansion of the universe.  Here the
parameter $M$ has the dimension of mass, and the scalar curvature is
defined as $R=g^{\mu\nu}g^{\alpha\beta}R_{\mu\alpha\nu\beta}$.

\begin{figure}[t]
  \centering
  \includegraphics[width=4.5in]{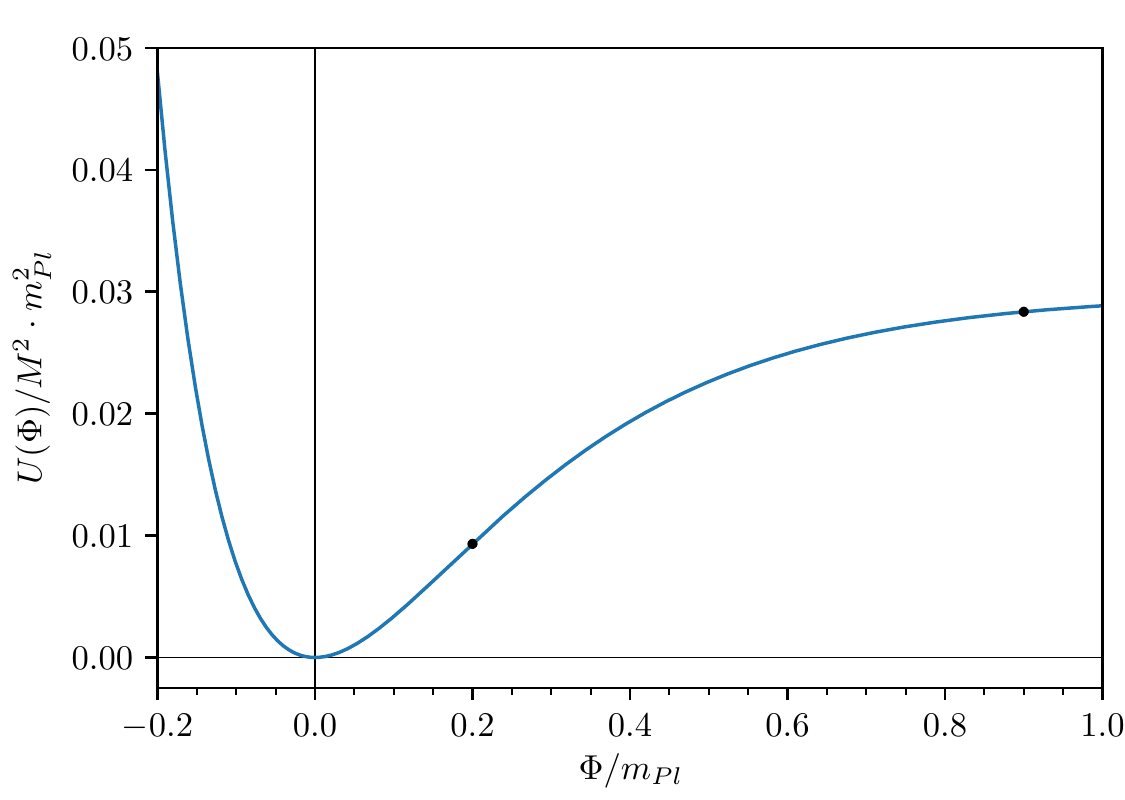}
  \caption{Inflaton potential $U(\Phi)$ in the $R^2$-inflation. Black
    dots correspond to $\Phi_i=0.9\, m_{Pl}$ and
    $\Phi_e=0.2\, m_{Pl}$.}
  \label{fig:R2inflation_U}
\end{figure}

Such $R^2$-gravity is similar to the general relativity with a scalar
field $\Phi$~\cite{Stelle:1978} with the potential (see
Fig.~\ref{fig:R2inflation_U})
\begin{equation}
  U(\Phi)=\frac{3M^2 m_{Pl}^2}{32\pi} \left(1-e^{-4\sqrt{\pi/3}\,\Phi/m_{Pl}} \right)^2.
\end{equation}

The slow-roll parameters are
\begin{align}
  \epsilon(\Phi)&=\frac{m_{Pl}^2}{16\pi}\left(\frac{U'(\Phi)}{U(\Phi)}\right)^2
                  =\frac{4}{3}\frac{1}{\left(e^{4\sqrt{\pi/3}\,\Phi/m_{Pl}}-1 \right)^2},\\
  \eta(\Phi)&=\frac{m_{Pl}^2}{8\pi}\frac{U''(\Phi)}{U(\Phi)}
              =-\frac{4}{3}\frac{e^{4\sqrt{\pi/3}\,\Phi/m_{Pl}}-2}{\left(e^{4\sqrt{\pi/3}\,\Phi/m_{Pl}}-1 \right)^2}.
\end{align}

One can see that $\epsilon(\Phi), |\eta(\Phi)| > 1$ for $\Phi < 0$,
therefore the slow-roll inflation takes place only if $\Phi>0$. The
inflation ends when $\epsilon(\Phi) \sim 1$ or $|\eta(\Phi)| \sim 1$,
so in our model we choose $\Phi_e=0.2\, m_{Pl}$.

The Hubble parameter is 
\begin{equation}
  \label{eq:H1}
  H(t)\simeq \sqrt{\frac{8\pi}{3m_{Pl}^2}U(\Phi(t))}=
  \frac{M}{2}\left(1-e^{-4\sqrt{\pi/3}\,\Phi(t)/m_{Pl}} \right).
\end{equation}

The amplitude of the scalar density perturbations is
\begin{equation}
  \Delta_R=\frac{3H^3}{2\pi |U'(\Phi)|}=
  \sqrt{\frac{3}{\pi}}\frac{M}{m_{Pl}}\sinh^2\left(2\sqrt{\frac{\pi}{3}}\frac{\Phi}{m_{Pl}} \right),
\end{equation}
here the value of $\Phi$ should be taken at the moment of 50--60
e-foldings before the end of inflation.  Using the measured magnitude
of $\Delta_R\simeq 5\cdot 10^{-5}$ one can calculate the value of mass
parameter, $M\simeq 2.6\cdot 10^{-6}\,m_{Pl}$.

We choose the time when the domain wall is
formed not at the very beginning of inflation,
$\Phi_i=\Phi(0)=0.9\, m_{Pl}$ (see Fig.~\ref{fig:R2inflation_U}).

Solution $\Phi(t)$ of the equation of motion,
$3H(t)\dot{\Phi}(t)\approx -U'(\Phi)$, during the slow-roll regime of
inflation can be found analytically,
\begin{equation}
  \label{phi(t)}
  \Phi(t)=\sqrt{\frac{3}{\pi}}\frac{m_{Pl}}{4}\log\left(e^{4\sqrt{\pi/3}\,\Phi_{i}/m_{Pl}}-\frac{2Mt}{3} \right).
\end{equation}
From this equation we can find the duration of the inflation: $t_e \approx 56.3\cdot M^{-1}$.

\begin{figure}[t]
  \centering
  \begin{subfigure}[b]{0.493\textwidth}
    \centering
    \includegraphics[width=3.0in]{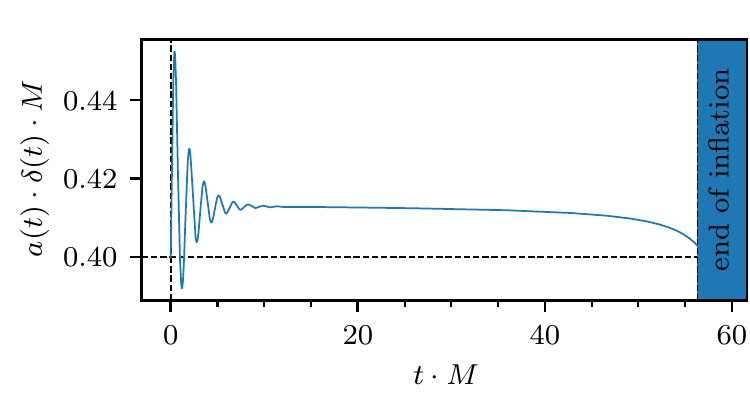}
    \caption{$\delta_0=0.4\cdot M^{-1}$}
    \label{fig:R2inflation=0.4}
  \end{subfigure}
  \begin{subfigure}[b]{0.493\textwidth}
    \centering
    \includegraphics[width=3.0in]{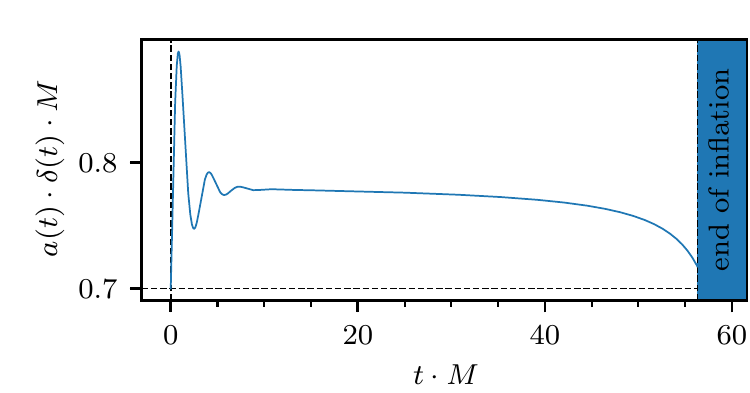}
    \caption{$\delta_0=0.7\cdot M^{-1}$}
    \label{fig:R2inflation=0.7}
  \end{subfigure}\\
  \begin{subfigure}[b]{0.493\textwidth}
    \centering
    \includegraphics[width=3.0in]{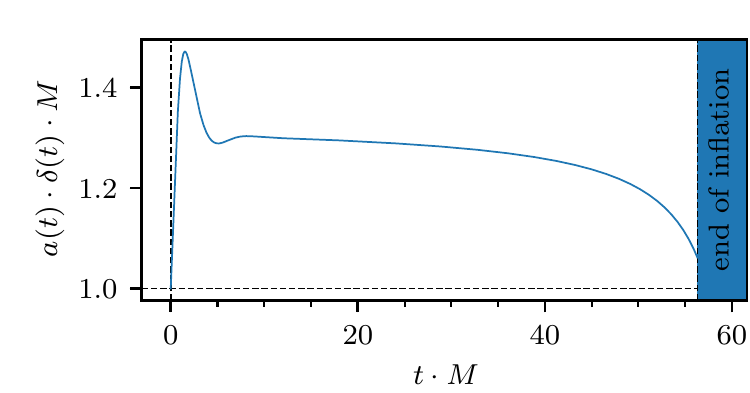}
    \caption{$\delta_0=1.0\cdot M^{-1}$}
    \label{fig:R2inflation=1.0}
  \end{subfigure}
  \begin{subfigure}[b]{0.493\textwidth}
    \centering
    \includegraphics[width=3.0in]{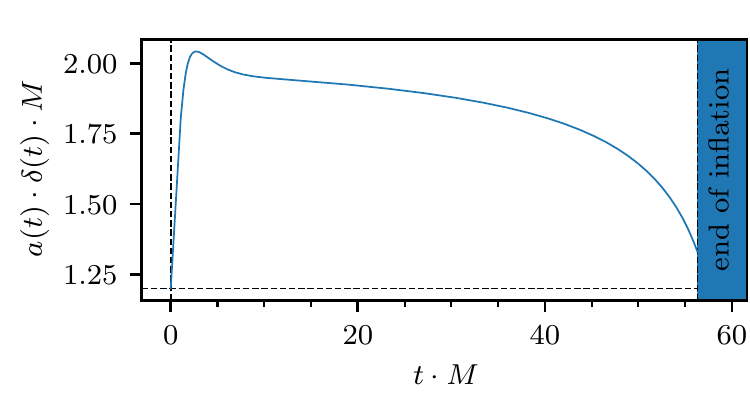}
    \caption{$\delta_0=1.2\cdot M^{-1}$}
    \label{fig:R2inflation=1.2}
  \end{subfigure}\\
  \begin{subfigure}[b]{0.493\textwidth}
    \centering
    \includegraphics[width=3.0in]{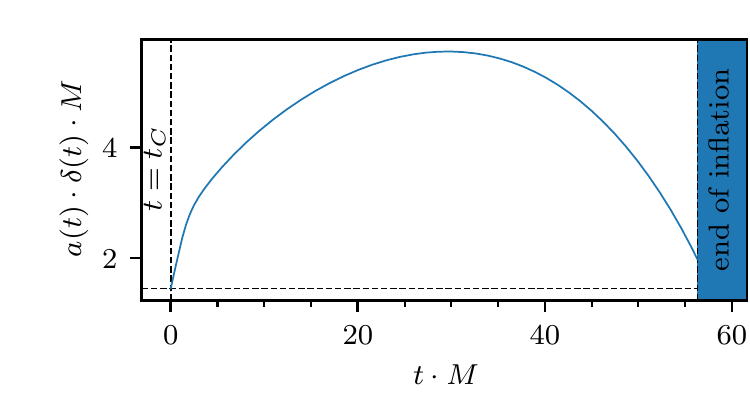}
    \caption{$\delta_0\approx1.4507\cdot M^{-1}$ $(t_{C}=0)$}
    \label{fig:R2inflation=1.4507}
  \end{subfigure}
  \begin{subfigure}[b]{0.493\textwidth}
    \centering
    \includegraphics[width=3.0in]{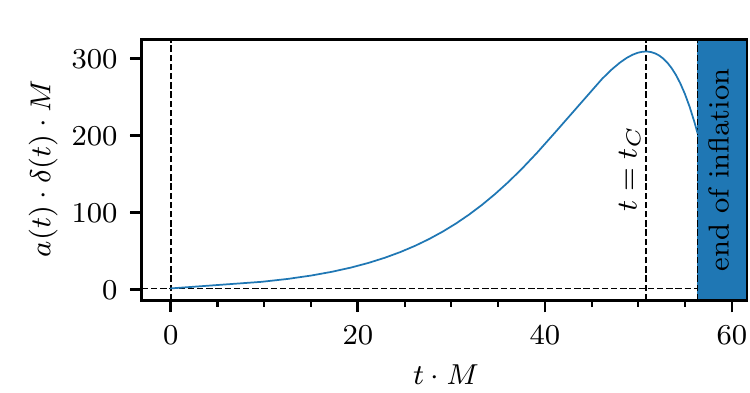}
    \caption{$\delta_0=1.7\cdot M^{-1}$}
    \label{fig:R2inflation=1.7}
  \end{subfigure}\\
  \begin{subfigure}[b]{0.493\textwidth}
    \centering
    \includegraphics[width=3.0in]{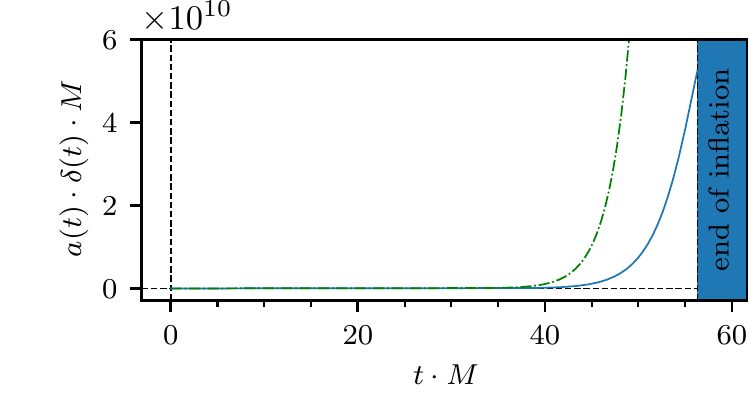}
    \caption{$\delta_0=5.0\cdot M^{-1}$}
    \label{fig:R2inflation=5.0}
  \end{subfigure}
  \begin{subfigure}[b]{0.493\textwidth}
    \centering
    \includegraphics[width=3.0in]{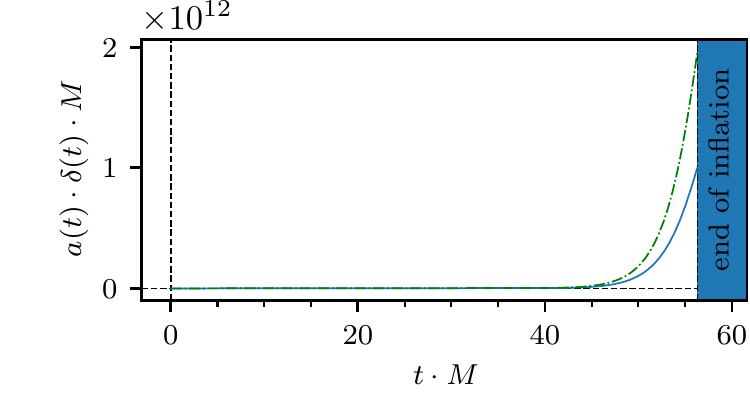}
    \caption{$\delta_0=10\cdot M^{-1}$}
    \label{fig:R2inflation=10}
  \end{subfigure}
  \caption{Time dependence of physical thickness of the wall,
    $a(t)\cdot\delta(t)$, for different values of the initial wall
    thickness, $\delta_0$, in the $R^2$-inflation.  Here $t$,
    $a(t)\delta(t)$ and $\delta_0$ are shown in $M^{-1}$ units.
    Dashed horizontal line corresponds to $\delta_0$.  The time at
    which the inflation ends is shown by vertical dashed line with the
    corresponding label. In Fig.~\ref{fig:R2inflation=1.4507} and
    \ref{fig:R2inflation=1.7} there is a vertical dashed line which
    corresponds to $t=t_C$. In Figs.~\ref{fig:R2inflation=5.0} and
    \ref{fig:R2inflation=10} there is a green (dash-dotted) line which
    corresponds to $a(t)\cdot\delta_0\cdot m$.}
  \label{fig:R2inflation_all}
\end{figure}

Substituting (\ref{phi(t)}) into (\ref{eq:H1}) one obtains
\begin{equation}
  \label{eq:R2_Ht}
  H(t)=\frac{M}{2}\left(1-\frac{1}{e^{4\sqrt{\pi/3}\,\Phi_{i}/m_{Pl}}-\frac{2Mt}{3}} \right),
\end{equation}
and correspondingly the scale factor is
\begin{equation}
  a(t)=a_0\cdot {\rm exp}\left(\int_0^t{H(t')dt'}\right)=
  a_0\cdot e^{Mt/2}\cdot \left(1-\frac{2Mt}{3}e^{-4\sqrt{\pi/3}\,\Phi_{i}/m_{Pl}} \right)^{3/4},
\end{equation}
in calculations we use $a_0=1$, therefore $a(0)=1$.

Since the Hubble parameter has the form~(\ref{eq:R2_Ht}), it is
convenient to use $M^{-1}$ units for $t$ and $\delta\left(t\right)$ in this
model, like we used $m^{-1}$ units in $\Phi^{2}$-model,
see~(\ref{eq:inf_eom}).

Now we can solve numerically the equation describing the domain wall
(\ref{eq-mot-f}).  The boundary conditions as usually are
(\ref{bound_cond1}) and we choose the initial configuration as the
domain wall with physical thickness $\delta_0$ and zero time
derivative (\ref{in_cond1}).

With the help of~(\ref{eq:H1}) we obtain the values of $\delta_0$
for which we have $t_C=0$ and $t_C=t_e$:
\begin{align}
  &t_C=0\text{ for
    }\delta_0=\frac{1}{H\left(0\right)\sqrt{2}}=\frac{\sqrt{2}}{1-e^{-4\sqrt{\pi/3}\,\Phi_i/m_{Pl}}}\cdot M^{-1}
    \approx 1.4507\cdot M^{-1},\\
  &t_C=t_e\text{ for
    }\delta_0=\frac{1}{H\left(t_e\right)\sqrt{2}}=
    \frac{\sqrt{2}}{1-e^{-4\sqrt{\pi/3}\,\Phi_e/m_{Pl}}}\cdot M^{-1}
    \approx 2.5300\cdot M^{-1}.
\end{align}

The results of numerical calculation for the time dependence of
physical thickness of the wall, $a(t)\delta(t)$, for different values of
parameter $\delta_0$ are presented in Fig.~\ref{fig:R2inflation_all}.

The results are similar to what we have seen in $\Phi^{2}$ and hilltop
models, i.e. the plots can be separated into three categories: very
thin walls, see
Figs.~\ref{fig:R2inflation=0.4}--\ref{fig:R2inflation=1.2}; medium
size walls, see
Figs.~\ref{fig:R2inflation=1.4507},~\ref{fig:R2inflation=1.7}; very
thick walls, see
Figs.~\ref{fig:R2inflation=5.0},~\ref{fig:R2inflation=10}. In the
latter case the thickness of the wall is growing almost with the size of
the universe. Summing up, we conclude that the particular model of
inflation is not that important, and in the case of fast expansion of
the universe the walls that are thick enough will be growing with the
scale factor.


\section{Evolution of thick domain walls in
  {\lowercase{\large $p=w\rho$}} universe}
\label{sec:w}

Now let us study how the domain wall evolves in an
postinflationary universe with the equation of state of matter $p=w\rho$,
where constant $w>-1$.  In such universe the scale factor increases as
some power of time,
\begin{equation}
  a(t)=\mathrm{const} \cdot t^\alpha, \hspace{3mm}
  \mathrm{where} \hspace{2mm}
  \alpha=\frac{2}{3(1+w)}>0,
  \label{eq:alpha}
\end{equation}
and the Hubble parameter decreases as inverse time,
\begin{equation}
  \label{eq:w_H}
  H(t)=\frac{\dot{a}}{a}=\frac{\alpha}{t}.
\end{equation}
The values $w=0$ $(\alpha=2/3)$ and $w=1/3$ $(\alpha=1/2)$ correspond
to the matter-dominated and radiation-dominated universe,
respectively.

After the substitution $\tau=t/\delta_0$, $\zeta=z/\delta_0$ into 
the equation of motion~(\ref{eq-mot-f}) we get
\begin{equation}
  \label{eq:eom}
  \frac{\partial^2 \tilde{f}}{\partial
    \tau^2}+\frac{3}{\sqrt{C\left(\tau\right)}}\frac{\partial \tilde{f}}{\partial
    \tau}-\frac{1}{\tilde{a}^2(\tau)}\frac{\partial^2\tilde{f}}{\partial\zeta^2}
  =2 \tilde{f} \left(1-\tilde{f}^2\right),
\end{equation}
where
$\tilde{f}(\zeta,\tau)=f(\zeta\cdot\delta_0,\tau\cdot\delta_0)$,
$\tilde{a}\left(\tau\right)=a\left(\tau\cdot\delta_0\right)$, and
\begin{equation}
  C\left(\tau\right)=
  \left(H\left(\tau\cdot\delta_0\right)\cdot\delta_0\right)^{-2}
  =H^{-2}\left(\tau\right).
  \label{eq:C}
\end{equation}
Here we used the explicit expression (\ref{eq:w_H}) for
$H\left(t\right)$, so the relation $C(\tau)=H^{-2}(\tau)$ is true only for $p=w\rho$ universe. 
For general dependence $H(t)$ the equality
$H(t)\delta_0=H\left(t/\delta_0\right)$ does not hold, so the
behaviour of thick and thin domain walls can be completely different
like it was in the case of de Sitter universe, $H={\rm const}$ (see
Section~\ref{sec:deSitter}). However, due to this feature of the
$p=w\rho$ universe there is no explicit dependence on $\delta_0$
in~(\ref{eq:eom}). This means that the equation of motion is the same
for different $\delta_0$ if $t$ and $z$ are measured in units of
$\delta_0$. Till the end of this section we will use $(t,z)$
notations considering $\delta_0=1$.

For the scale factor we choose the following form:
$a(t)=a_0\cdot\left({t}/{t_{i}}\right)^{\alpha}$ with the constant
$a_0=1$, so $a\left(t_{i}\right)=1$ which means that the initial
coordinate and physical distances are the same.

\begin{figure}[t]
  \centering
  \begin{subfigure}[b]{\textwidth}
    \includegraphics[width=6.08in]{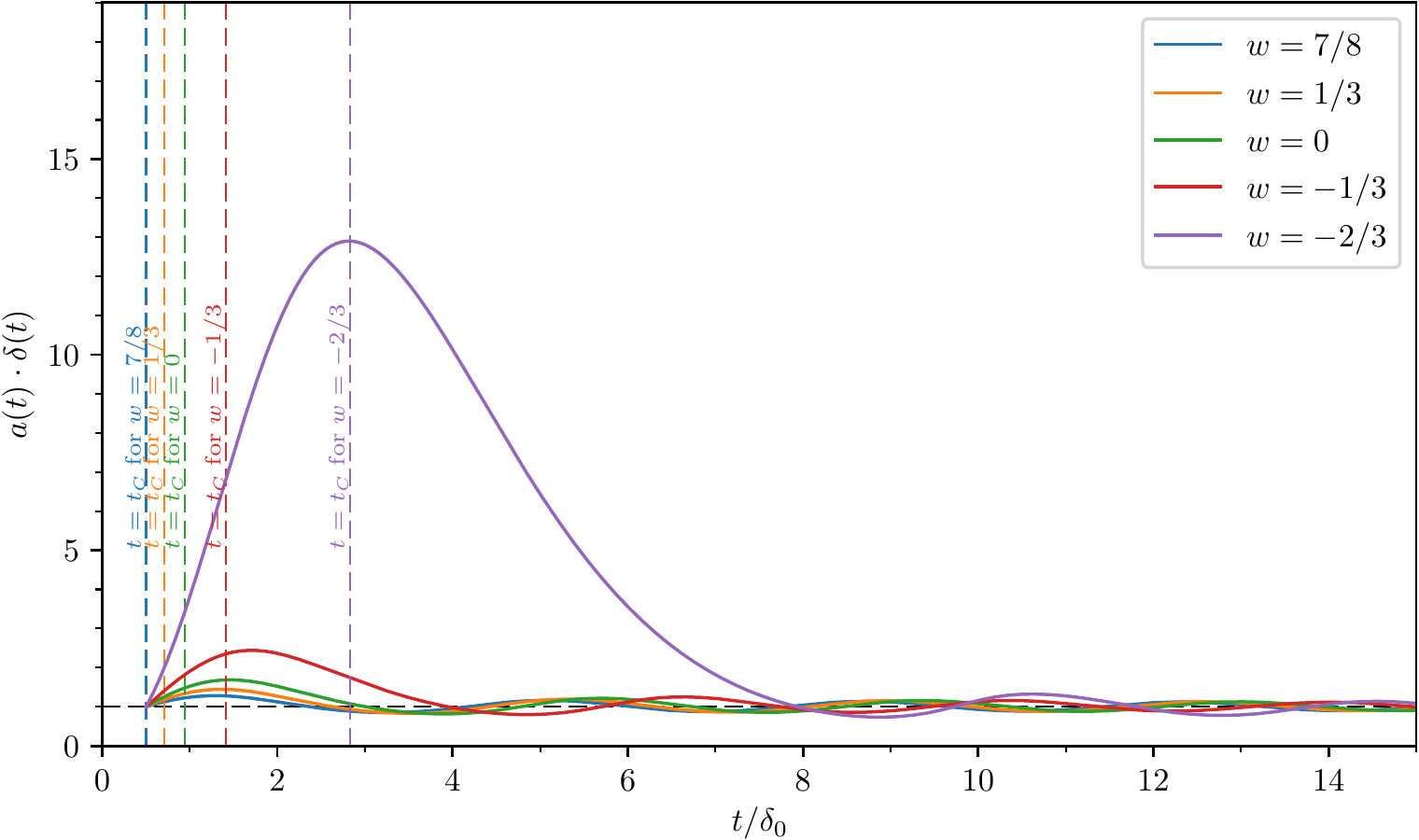}
    \caption{$t_{i}/\delta_0=0.5$}
    \label{fig:ti05}
  \end{subfigure}\\
  \begin{subfigure}[b]{\textwidth}
    \includegraphics[width=6.08in]{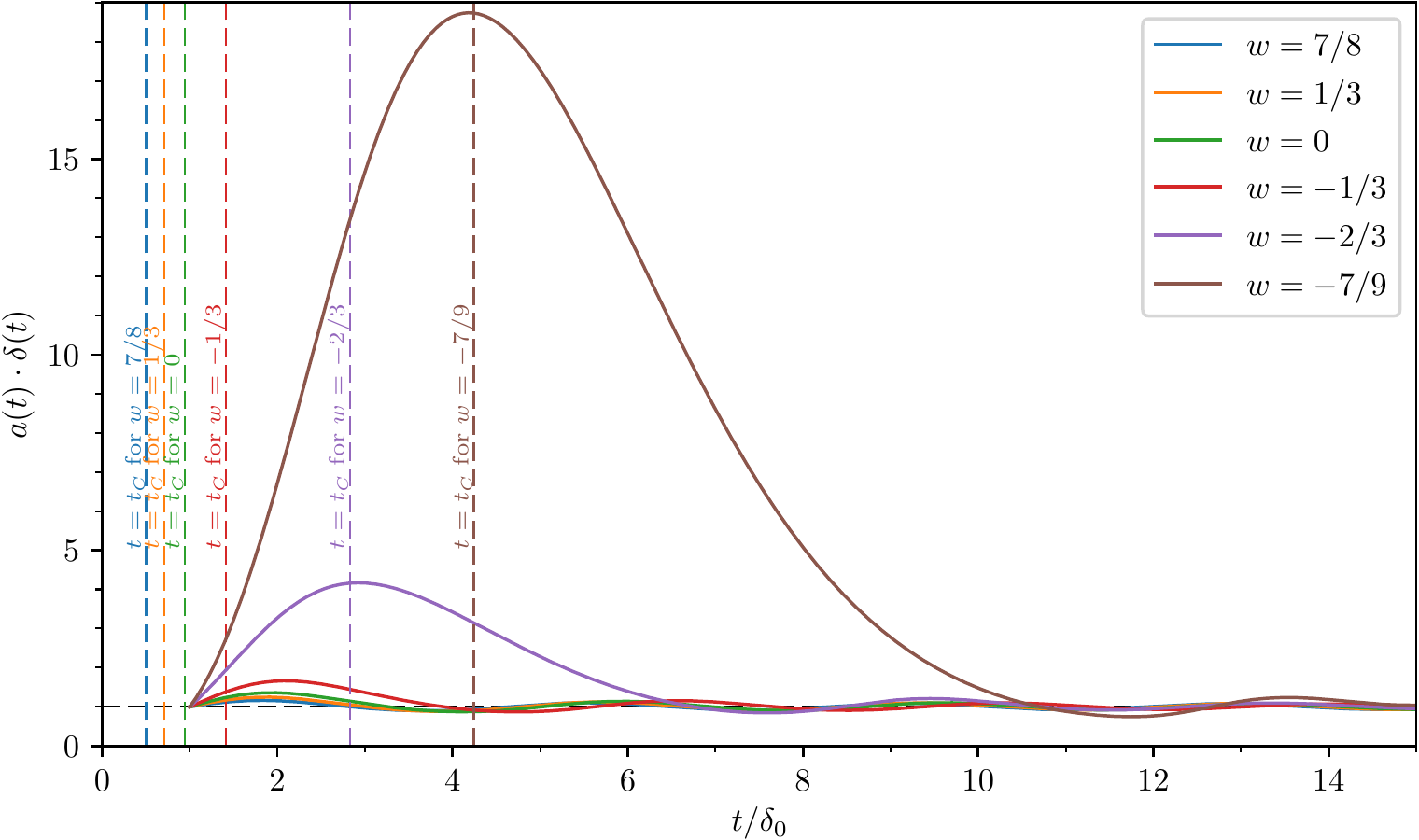}
    \caption{$t_{i}/\delta_0=1.0$}
    \label{fig:ti10}
  \end{subfigure}
  \caption{Time dependence of the physical thickness of the wall,
    $a(t)\delta(t)$, for different values of parameter $w$.  Dashed
    horizontal line corresponds to $\delta_0$. Vertical dashed lines
    correspond to the moment $t_{C}$ at which $C\left(t_{C}\right)=2$
    (the correspondence between color and $w$ is the same).}
      \label{fig:w_all}
\end{figure}

The results of numerical calculation for the time dependence of
physical thickness of the wall, $a(t)\delta(t)$, in radiation-dominated
and matter-dominated universe are presented in Fig.~\ref{fig:w_all}
along with the plots for other values of~$w$.
In Figs.~\ref{fig:ti05} and \ref{fig:ti10} the initial time is chosen
to be $t_{i}/\delta_0=0.5$ and $t_{i}/\delta_0=1.0$,
respectively. We can see that these two plots look very much alike but
the curves for $t_{i}/\delta_0=0.5$ look similar to the curves for
$t_{i}/\delta_0=1.0$ with another value of $w$ (i.e. curve with
$w=-2/3$ and $t_{i}/\delta_0=0.5$ reaches approximately the same
maximum value as the curve with $w=-7/9$ and
$t_{i}/\delta_0=1.0$). Let us explain this.
  
\begin{figure}[t]
  \centering
  \begin{subfigure}[b]{\textwidth}
    \includegraphics[width=6.08in]{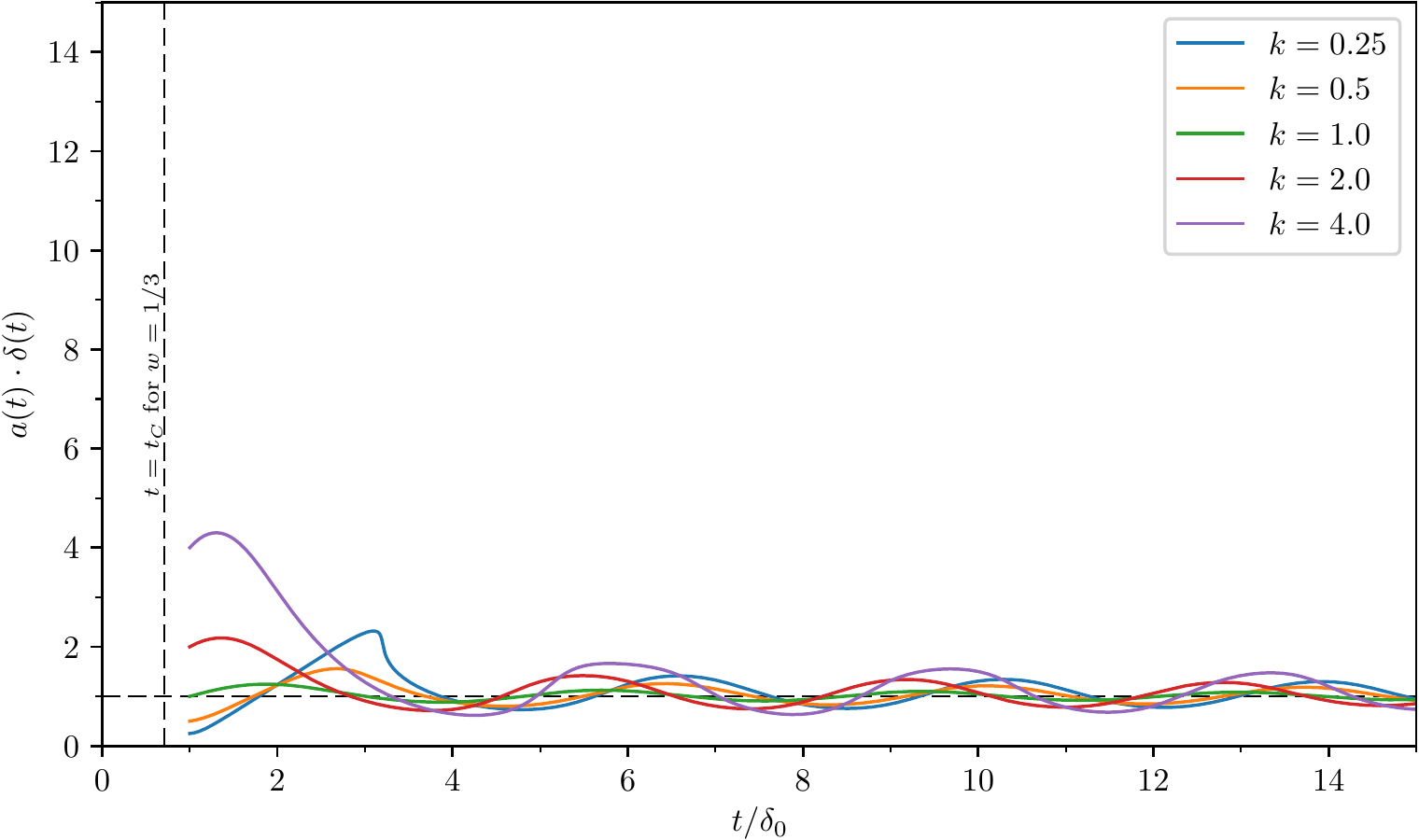}
    \caption{$w=\frac{1}{3}$}
    \label{fig:stretch_1_3}
  \end{subfigure}
  \begin{subfigure}[b]{\textwidth}
    \includegraphics[width=6.08in]{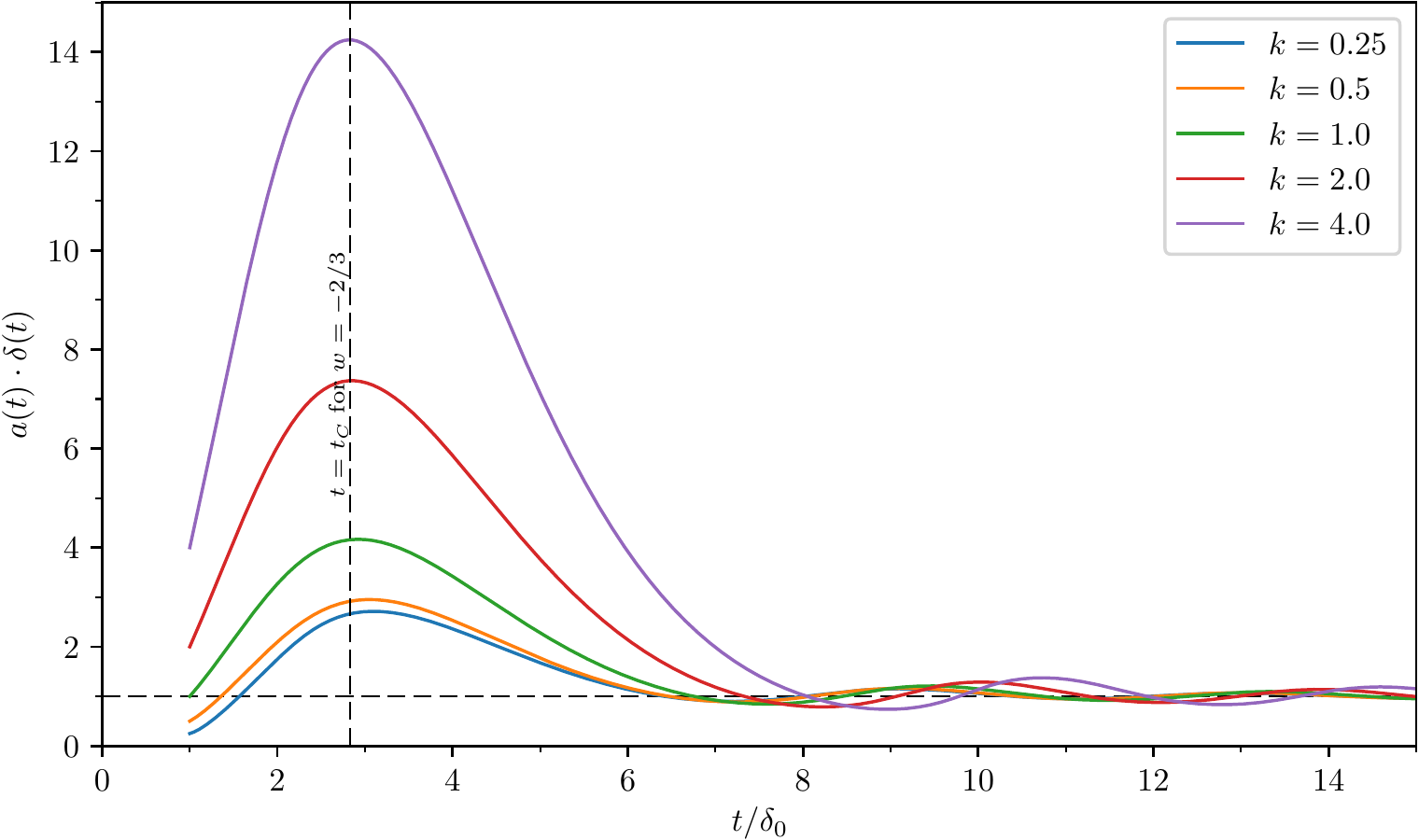}
    \caption{$w=-\frac{2}{3}$}
    \label{fig:stretch_-2_3}
  \end{subfigure}
  \caption{Time dependence of the physical thickness of the wall,
    $a(t)\delta(t)$, for different values of stretch parameter $k$.
    Dashed horizontal line corresponds to $\delta_0$. For both plots
    $t_{i}/\delta_0=1.0$. Vertical dashed lines correspond to the
    moment $t_{C}$ at which $C\left(t_{C}\right)=2$.}
  \label{fig:stretch}
\end{figure}

The evolution of the domain wall is basically determined by the parameter
$C\left(t\right)$, which increases in the $p=w\rho$ universe as
\begin{equation}
  C(t)=\frac{1}{(H(t)\delta_0)^2} =
  \frac{t^2}{(\alpha\delta_0)^2}\propto t^2.
\end{equation}

Since $C=2$ is the critical value, let us introduce the time $t_C$ at
which $C(t_{C})=2$. In $p=w\rho$ universe
\begin{equation}
  \frac{t_C}{\delta_0}=\sqrt{2}\alpha.
  \label{eq:tC}
\end{equation}

With the help of (\ref{eq:alpha}) we obtain that
$t_{C}>t_{i}$ for
\begin{equation}
  w<\frac{2\sqrt{2}}{3}\frac{\delta_0}{t_{i}}-1.
  \label{eq:tC_ti}
\end{equation}

Let us consider $t_{i}/\delta_0=1.0$ (see
Fig.~\ref{fig:ti10}). Using (\ref{eq:tC_ti}) we get that $t_{C}>t_{i}$
for $w<2\sqrt{2}/3-1\approx -0.057$. In this case $C(t)$ is
larger than $2$ during all the time of the wall evolution for $w=1/3$
and $w=0$. If we consider the value of parameter $w$ which is
different from $1/3$ or $0$, e.g. $w=-1/3$
$(\alpha=1,\ \dot{a}(t)=const)$, $w=-2/3$
$(\alpha=2,\ \ddot{a}(t)=const)$, $w=-7/9$
$(\alpha=3,\ \dddot{a}(t)=const)$, then at the initial moment
parameter $C$ can be smaller than 2. In Fig.~\ref{fig:ti10} one can
see that in this case, while $C<2$, the physical thickness of the wall,
$a(t)\delta(t)$, increases rapidly. Moreover, when the parameter $w$
approaches $(-1)$, the thickness of the wall is growing much faster.
However, the parameter $C(t)$ increases with time and eventually
becomes larger than 2. After that the wall thickness starts to decrease.

In case of $t_i/\delta_0=0.5$ (see Fig.~\ref{fig:ti05}), for the
same $w$ the wall thickness is growing for a longer period of time,
$t_C-t_i\,$, since $t_C$ does not depend on $t_i$. We also can
say that with different $t_i$ the period of growth is the same for
different values of $w$. This explains why curves 
corresponding to different values of $w$ do look alike in Figs.~\ref{fig:ti05} and \ref{fig:ti10}.

As for behaviour at $t\to\infty$, one sees in Fig.~\ref{fig:w_all}
that the physical thickness oscillates with slowly decreasing amplitude around
the value $\delta_0$.  Therefore, when $t\to\infty$ the field
configuration tends to
\begin{equation}
  \label{final-kink}
  f(z,t)=\tanh{\frac{z\cdot a(t)}{\delta_0}}.
\end{equation}
This asymptotic behaviour can be understood from \eqref{eq-mot-f}. If
the first two terms in the l.h.s. of \eqref{eq-mot-f} can be
neglected, then this equation becomes kink-type one, and has the
solution \eqref{final-kink}.

In Fig.~\ref{fig:w_all} we present the domain walls with the initial
physical thickness of the ones in stationary universe ($H=0$).  By doing
that, we separate the influence of cosmological expansion from
the natural shrinking/expanding of the wall when it approaches
stationary solution. However, we can consider the evolution of the
wall thickness from different initial configurations. In
Fig.~\ref{fig:stretch} we can see the evolution of the wall thickness
from the configurations stretched along $z$ by factor $k$:
\begin{equation}
  \label{in_cond_k}
  f(z,t_i)=\tanh{\frac{z\cdot a\left(t_{i}\right)}{k\cdot\delta_0}}.
\end{equation}
For $w=1/3$ (so $t_{C}<t_{i}=1\cdot\delta_0$) all initial configurations quickly
approach their common stationary solution. On the other hand, for
$w=-2/3$ (so $t_{C}>t_{i}=1\cdot\delta_0$) there is a period of growth for all
initial configurations though it is worth noting that this growth
depends on $k$.

To conclude this section, we note that in $p=w\rho$ universe it is
difficult to obtain domain walls with cosmologically large thickness ($w$
should be really close to $-1$ for that). Such domain walls can exist
at the beginning of $p=w\rho$ stage. However, in such universe the
parameter $C(t)$ increases, so at some moment the wall thickness starts to
shrink and eventually goes to the constant value, $\delta_0$, which is
microscopically small.


\section{Conclusions}
\label{sec:conclusion}

We have found that at inflationary epoch the thickness of the domain walls exponentially rises 
when the parameter $C(t)$ is smaller than the critical value $C= 2$. 
Therefore, the wall thickness might rise up to cosmologically large
scales if $C(t)$ remained smaller than 2 till the end of inflation. 
However, when inflation is over and the expansion turns into the power
law regime with $H$ decreasing with time as $1/t$, the parameter
$C(t) \sim 1/H^2(t) \sim t^2$ at some moment becomes larger than 2,
and the wall started to shrink.

  Our original scenario~\cite{dgrt} spans throughout the
  inflation and ends in the beginning of the reheating stage where the
  baryogenesis is supposed to take place. That is why we are
  interested in the wall evolution during all the inflation and at the
  beginning of the $p=w\rho$ stage. The main result of our
  calculations is that there exists sufficiently wide range of the
  parameters for which the domain wall remains astronomically thick during
  the baryogenesis epoch. This ensures large separation between
  domains of matter and antimatter and prevents from catastrophic
  annihilation. This result allows to determine the range of the
  parameters of realistic baryogenesis models with safe separation
  between matter and antimatter.

An efficient baryogenesis could start only after inflation,
and moreover in our model \cite{dgrt} 
it should take place before the field $\varphi$ rolled
    down to zero (here $\varphi$ is the field which made the
    wall). This can easily be achieved with a proper choice
    of the model parameters, as is shown in~\cite{dgrt}. The
    particular mechanism of inflation is not essential for our results.
    The only thing which we need is the exponential cosmological
    expansion.

We have considered here the evolution of a single domain wall,
rather than that of a whole network of walls.
Indeed, knowledge of the wall speeds and interaction or annihilation rates would be important 
only if the domain size is smaller than the cosmological horizon.
Normally the size of the domain is much larger
  than the thickness of the wall. So the size of the domain can be easily
  much larger than the horizon at the baryogenesis epoch. Soon after
  it the domain walls disappeared completely and their evolution became not
  essential. Here is a substantial difference between the usually
  considered case of the non-destructible domain walls and our model~\cite{dgrt}
  with disappearing domain walls. 
  
  The fact that the wall thickness was much larger than the horizon also means that it is
  much larger than the diffusion length of the nucleons or quarks and
  the effects of their annihilation are not essential prior the domain
  wall disappearance.

As is argued in ref.~\cite{CdRG}, the size of the domain walls cannot
be much smaller than $\sim 10$ Mpc, since otherwise the annihilation
would be too strong. On the other hand, it cannot be noticeably larger
that the same 10 Mpc to avoid too large angular fluctuations of CMB
above the diffusion (Silk) damping scale. Based on this result the
authors of~\cite{CdRG} have concluded that the nearest antimatter
domain should be at the distance of a few Gigaparsec away. This result
is true in baryosymmetric universe. However, if the cosmological
fraction of antibaryons is much smaller or larger than that of baryons
the limit would be proportionally weakened.

Note in this connection that the spontaneous $CP$ violation does not
necessarily lead to equal cosmological densities of matter and
antimatter but to an excess or deficit of one with respect to the
other, see e.g. the lectures~\cite{AD-CP-cosm}.

Another possible concern about the result of ref.~\cite{CdRG} is that
it has been derived in the approximation of uniform CMB temperature.
Different rates of the cosmological expansion near the wall and far
from it could lead to different temperatures at the scales below the
diffusion damping. This effect might accelerate or slow down the
proton or antiprotons diffusion into "hostile" antimatter or matter
environment and respectively either amplify or inhibit the
annihilation. This problem is under investigation now.

As rather far fetched but exciting idea let us mention the possibility
that the maybe-observed deficit of baryons in the 300 Mpc local
universe~\cite{under-B} can be explained in the
model of mildly inhomogeneous baryogenesis with thick domain
walls.


\section*{Acknowledgements}
We acknowledge support of the RSF Grant No. 16-12-10037.
SG is also supported under the grants RFBR
No.~16-32-60115, 16-32-00241, 16-02-00342. 
In addition, SG is grateful to Dynasty Foundation for support.

\bibliographystyle{apsrev4-1}
\bibliography{references}

\end{document}